\documentclass[aps,prl,showpacs,preprintnumbers,amsmath,amssymb,twocolumn,letterpaper,superscriptaddress]{revtex4-1}
\usepackage{amssymb}
\usepackage{ifpdf}
\usepackage[utf8]{inputenc}
\usepackage{graphicx}
\usepackage{times}
\usepackage{dcolumn}
\usepackage{textcomp}
\usepackage{amsmath}
\usepackage{array}
\usepackage{float}
\usepackage{bibunits}

\makeatletter
\newcommand*{\balancecolsandclearpage}{%
  \close@column@grid
  \clearpage
  \twocolumngrid
}
\makeatother

\def\UCFG{UCo$_{1-x}$Fe$_x$Ge}
\def\UCG{UCoGe}

\begin{document}

\preprint{}

\title{Quantum critical scaling in the disordered itinerant ferromagnet UCo$_{1-x}$Fe$_x$Ge}

\author{K. Huang}
\affiliation{Condensed Matter and Magnet Science, Los Alamos National Laboratory, Los Alamos, New Mexico 87545, USA}
\affiliation{Department of Physics, University of California, San Diego, La Jolla, California 92093, USA}
\author{S. Eley}
\affiliation{Condensed Matter and Magnet Science, Los Alamos National Laboratory, Los Alamos, New Mexico 87545, USA}
\author{P. F. S. Rosa}
\affiliation{Condensed Matter and Magnet Science, Los Alamos National Laboratory, Los Alamos, New Mexico 87545, USA}
\author{L. Civale}
\affiliation{Condensed Matter and Magnet Science, Los Alamos National Laboratory, Los Alamos, New Mexico 87545, USA}
\author{E. D. Bauer}
\affiliation{Condensed Matter and Magnet Science, Los Alamos National Laboratory, Los Alamos, New Mexico 87545, USA}
\author{R. E. Baumbach}
\altaffiliation[Current address: ]{Condensed Matter Group, National High Magnetic Field Laboratory, Florida State University, Tallahassee, Florida 32310, USA}
\affiliation{Condensed Matter and Magnet Science, Los Alamos National Laboratory, Los Alamos, New Mexico 87545, USA}
\author{M. B. Maple}
\affiliation{Department of Physics, University of California, San Diego, La Jolla, California 92093, USA}
\author{M. Janoschek}
\email[Corresponding Author: ]{mjanoschek@lanl.gov}
\affiliation{Condensed Matter and Magnet Science, Los Alamos National Laboratory, Los Alamos, New Mexico 87545, USA}

\date{\today}

\begin{abstract}
Belitz-Kirkpatrick-Vojta (BKV) theory shows in excellent agreement with experiment that ferromagnetic quantum phase transitions (QPTs) in clean metals are generally first-order due to the coupling of the magnetization to electronic soft modes, in contrast to the classical analogue that is an archetypical second-order phase transition. For disordered metals BKV theory predicts that the second order nature of the QPT is restored because the electronic soft modes change their nature from ballistic to diffusive. Our low-temperature magnetization study identifies the ferromagnetic QPT in the disordered metal \UCFG\ as the first clear example that exhibits the associated critical exponents predicted by BKV theory.
\end{abstract}

\maketitle


\begin{bibunit}

Quantum phase transitions (QPTs) have been a topic of intense research efforts for several decades \cite{Loehneysen07, Gegenwart08, Sachdev:2011, Brando:2016}. Here the earliest theory of a QPT was provided by Stoner in 1938 for itinerant ferromagnets \cite{Stoner:1938}. Due to the exotic behavior frequently observed in the vicinity of ferromagnetic QPTs in metals, such as unconventional spin-triplet superconductivity~\cite{Pfleiderer09}, partial magnetic order \cite{Pfleiderer:2004}, and topological non-Fermi liquid behavior \cite{Ritz:2013}, their theoretical understanding is at the origin of modern solid state physics. In their seminal work, Hertz and later Millis predicted that ferromagnetic QPTs are continuous, or second-order, and calculated the associated critical exponents \cite{Hertz:1976, Millis:1993}. However, at the turn of the century, an extension of the Hertz-Millis theory by Belitz, Kirkpatrick and Votja (BKV) demonstrated in remarkable agreement with experiments that a QPT in two and three dimensions from a paramagnetic to homogeneous ferromagnetic state is generically discontinuous (or first-order) provided that the underlying metal is sufficiently clean \cite{Belitz:1999}. The responsible mechanism is the coupling of the magnetization to electronic soft modes that universally exist in metals, which in turn leads to a fluctuation-induced first-order transition. We note that fluctuation-induced first-order transitions are  broadly important in solid-state physics and even beyond~\cite{Janoschek:2013}.

BKV theory further reveals the existence of a tricritical point that separates a line of first-order transitions at low temperature from a line of second order transitions at higher temperatures when the non-thermal control parameter $x$ that provides access to the QPT is varied. Including the effects of an external magnetic field $H$ in the BKV calculations generates tricritical wings that emerge from the tricritical point as a function of magnetic field \cite{Belitz:2005}. The resulting unique temperature $T$ vs $x$ and $H$ phase diagram has been observed in experiments on many clean ferromagnetic metals, making BKV theory one of the most successful theories of QPTs \cite{Brando:2016}.

Because in a large number of itinerant ferromagnets the QPT may be accessed via chemical substitution, and some materials show incipient disorder, considering the effect of disorder on the nature of a ferromagnetic QPT is crucial. If the disorder is sufficiently strong, the nature of the electronic soft modes changes from ballistic to diffusive. This slowing-down of the itinerant electrons promotes ferromagnetism. According to BKV theory, this results in the suppression of the tricritical point to zero temperature and the resulting QPT is second-order~\cite{Belitz:2001a,Belitz:2001b, Kirkpatrick:1996}, with critical exponents that suggest the transition is even more continuous than in Hertz-Millis theory \cite{Brando:2016, Belitz:2015}. Although most disordered ferromagnetic metals exhibit second-order QPTs, the critical exponents predicted by BKV theory (reviewed below) have never been observed consistently.

In this Letter, we demonstrate that the critical behavior observed at a ferromagnetic QPT in \UCG\ that is accessed via chemical substitution of Co with Fe is in excellent agreement with BKV theory. \UCG\ orders ferromagnetically below a Curie temperature $T_{\rm{C}} = 3$~K and coexists with unconventional superconductivity below $T_{\rm{S}} = 0.8$~K~\cite{Huy:2007}.  Superconductivity in \UCFG\ is only observed for Fe concentrations $x\leq 0.025$ \cite{Huang:2013}. In contrast, $T_{\rm{C}}$ first increases to the maximum $T_{\rm{C}}\approx9$~K  at $x = 0.075 - 0.1$, and then smoothly decreases to zero temperature at $x_{cr} = 0.23$ consistent with a second-order QPT at $x_{cr}$ \cite{Huang:2013}. The observed increased values of the residual resistivity ($\rho_0(x_{cr})\approx$~420~$\mu\Omega$cm)\cite{Huang:2013} suggest a significant amount of disorder making \UCFG\ an ideal candidate to look for the critical exponents predicted by BKV theory for disordered metals. In our previous study, the magnetization near the QPT was found to scale as  $M$($T = 2$~K , $H$)~$\propto H^{1/\delta}$ as function of $H$. Here the corresponding critical exponent was determined to be $\delta\approx$~3/2 in agreement with BKV theory for a disordered QPT. Because strictly speaking the exponent $\delta$ takes on the value associated with the QPT only for $T=0$, this has motivated our present study of $M$($T$, $H$) down to much lower temperatures. Our results show that near the disordered ferromagnetic QPT in \UCFG, all critical exponents may be accurately determined from $M$ and agree quantitatively with BKV theory.

Samples of \UCFG\ were synthesized using a custom built single-arc furnace using a water-cooled copper hearth in argon atmosphere with a zirconium getter. The starting materials of U (99.9$\%$), Co pieces (99.99$\%$), Fe pieces, and Ge pieces (99.9999+$\%$), were weighed stoichiometrically, arc melted, flipped over and remelted five times to ensure chemical homogeneity. Chemical analysis of all samples was carried out using a commercial scanning electron microscope (FEI Inspect F) equipped with a energy dispersive spectroscopy (EDS) microprobe. The EDS analysis (see supplemental material \cite{Supp}) shows that they are indeed chemically homogeneous, where in particular the nominal Fe concentration $x$ agrees with the Fe concentration $x_{meas}$ within the error bar. Therefore, we use the nominal Fe concentration $x$ throughout the text. We note that polycrystalline samples were chosen purposefully to obtain the most reliable data. Specifically, the magnetic properties of \UCFG\ near the QPT are extremely sensitive to the Fe concentration as shown below. Single crystals of \UCFG\ are grown via the Czochralski method, which typically leads to concentration gradients that would be detrimental for the determination of critical exponents. Finally, it has been demonstrated that the use of polycrystalline samples does not affect the ability to reliably determine scaling exponents of phase transitions \cite{Fuchs:2014, Huang:2015}.

\begin{figure}[t]
\begin{center}
    \includegraphics[width=0.9\columnwidth]{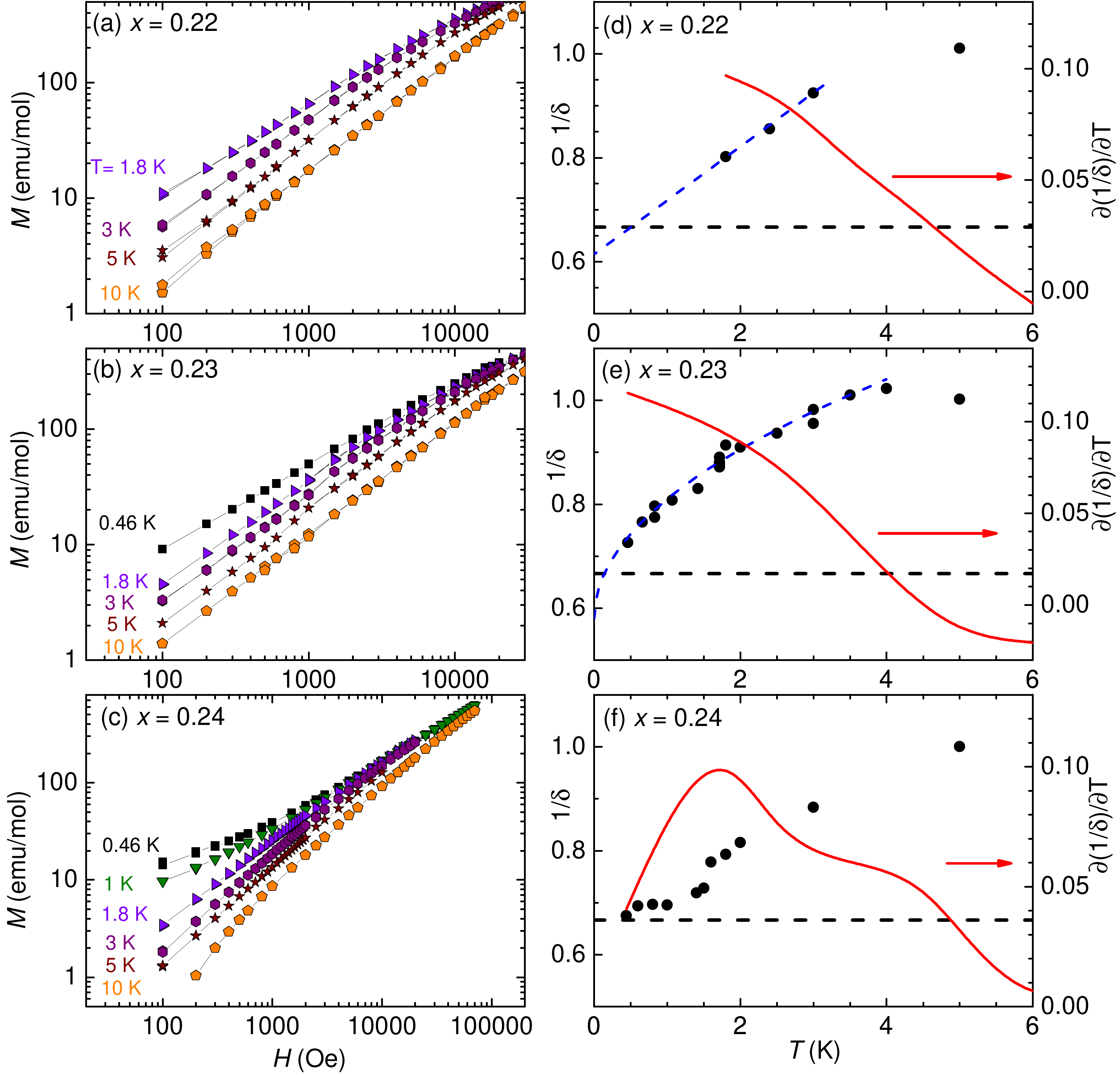}
    \caption{(Color online) Isotherms of the magnetization $M$($T$, $H$) of \UCFG\ as function of magnetic field $H$ for Fe concentrations $x = 0.22$ (a),  0.23 (b) and 0.24 (c) displayed in a log-log plot. Because  $M$($T$ , $H$)~$\propto H^{1/\delta}$ the slope of each curves describes $1/\delta$. In panels (d)-(f), the temperature-dependence of $1/\delta$ is shown for each concentration.  Because  for $H\leqq0.1$ T scaling is not expected due to domain effects, the corresponding data were omitted for determination of $1/\delta$~\cite{Huang:2015}. The blue dashed lines in (d)-(e) are guides to the eye. The horizontal dashed black line denotes $\delta=3/2$. The red solid curves is the partial derivative $\partial(1/\delta)/\partial T$ with respect to $T$.}
    \label{fig:delta}
\end{center}
\end{figure}

All magnetization $M$($T$, $H$) data presented here were obtained in a Quantum Design magnetic property measurement system (MPMS) with a $^3$He insert, reaching temperatures $T$ from 300 K down to 460 mK in fields up to 7 Tesla. Fig.~\ref{fig:delta} shows isotherms of the magnetization for various $T$ and $x = 0.22$ (a),  0.23 (b) and 0.24 (c). The data are displayed in a log-log plot so that the slope of each curve corresponds to $1/\delta$. We note that for the determination of $1/\delta$, the data for $H\leqq0.1$ T, where scaling is not expected because of domain effects, were omitted \cite{Huang:2015}. The resulting temperature dependence of $1/\delta$ for each of the three concentrations is shown in Fig.~\ref{fig:delta} (d)-(f). Inspecting Fig.~\ref{fig:delta}(f) for $x=0.24$, it is clear that $1/\delta$ saturates at 2/3 for $T \longrightarrow 0$, in excellent agreement with theory. For $x=0.22$ and $0.23$ no saturation is observed and the value of $1/\delta$ for $T \longrightarrow 0$ is more challenging to estimate.

Before continuing the discussion of our results, it is useful to recall the critical exponents that can be determined from magnetization data and their values as calculated via BKV theory for a three-dimensional, ferromagnetic QPT in a metal with significant disorder \cite{Brando:2016, Belitz:2015}. They are summarized in table~\ref{tab:critical}. Two regimes have to be considered: (a) the so-called asymptotic regime that should only exist in a narrow region near the QPT, and (b) the pre-asymptotic region that describes the critical exponents further away from the QPT \cite{Kirkpatrick:2014}. For $\delta$ the asymptotic and pre-asymptotic values are $3/2$ and $11/6$, respectively. This suggest that the $x=0.24$ is directly in the vicinity of the QPT and thus shows asymptotic behavior. In contrast, for $x=0.22$ and $0.23$ pre-asymptotic scaling is expected, and indeed provides an excellent description of our data as shown in detail below. The absence of low-temperature saturation of $1/\delta$ for $x=0.22$ and $0.23$ is explained by the fact that critical scaling is typically only observed to higher temperatures close to the QPT~\cite{Loehneysen07}. Therefore, measurements to lower temperatures than accessible in our experiments are required to determine $\delta$ unambiguously. Nevertheless, inspection of the partial derivative $\partial(1/\delta)/\partial T$ (Fig.~\ref{fig:delta} (d)-(f)) demonstrates that the slope of $1/\delta$ is finite for $T \longrightarrow 0$ (cf. for $x=0.24$, $\partial(1/\delta)/\partial T\longrightarrow 0$), suggesting that $1/\delta<2/3$ in the pre-asymptotic regime in agreement with BKV theory.

\begin{table}[t]
\caption{The critical exponents for a ferromagnetic second-order quantum phase transition for a "dirty" itinerant ferromagnet in three dimensions according to the theory by Belitz, Kirkpatrick and Votja (BKV)~\cite{Brando:2016, Belitz:2015} are provided for both the (a) asymptotic and (b) pre-asymptotic regimes. The (c) column denotes the corresponding values for an unstable Hertz type fixed point in three dimensions in the dirty limit.}\label{tab:critical}
\begin{tabular}{cccc}
\hline
Critical exponent & (a) Asymptotic & (b) Pre-asymptotic & (c) Hertz (dirty)\\
\hline\hline
$\delta$ & 3/2 & 11/6 & 3\\
$\beta_T$ & 1 & 3/4 & 5/8\\
$\gamma_T$ & 1/2 & 5/8 & 5/4\\
$\nu$ & 1 & 3/5 & 1/2\\
$z_m$ & 2 & 8/3 & 8/5\\
\hline
\end{tabular}
\end{table}

\begin{figure}[h!]
\begin{center}
    \includegraphics[width=0.95\columnwidth]{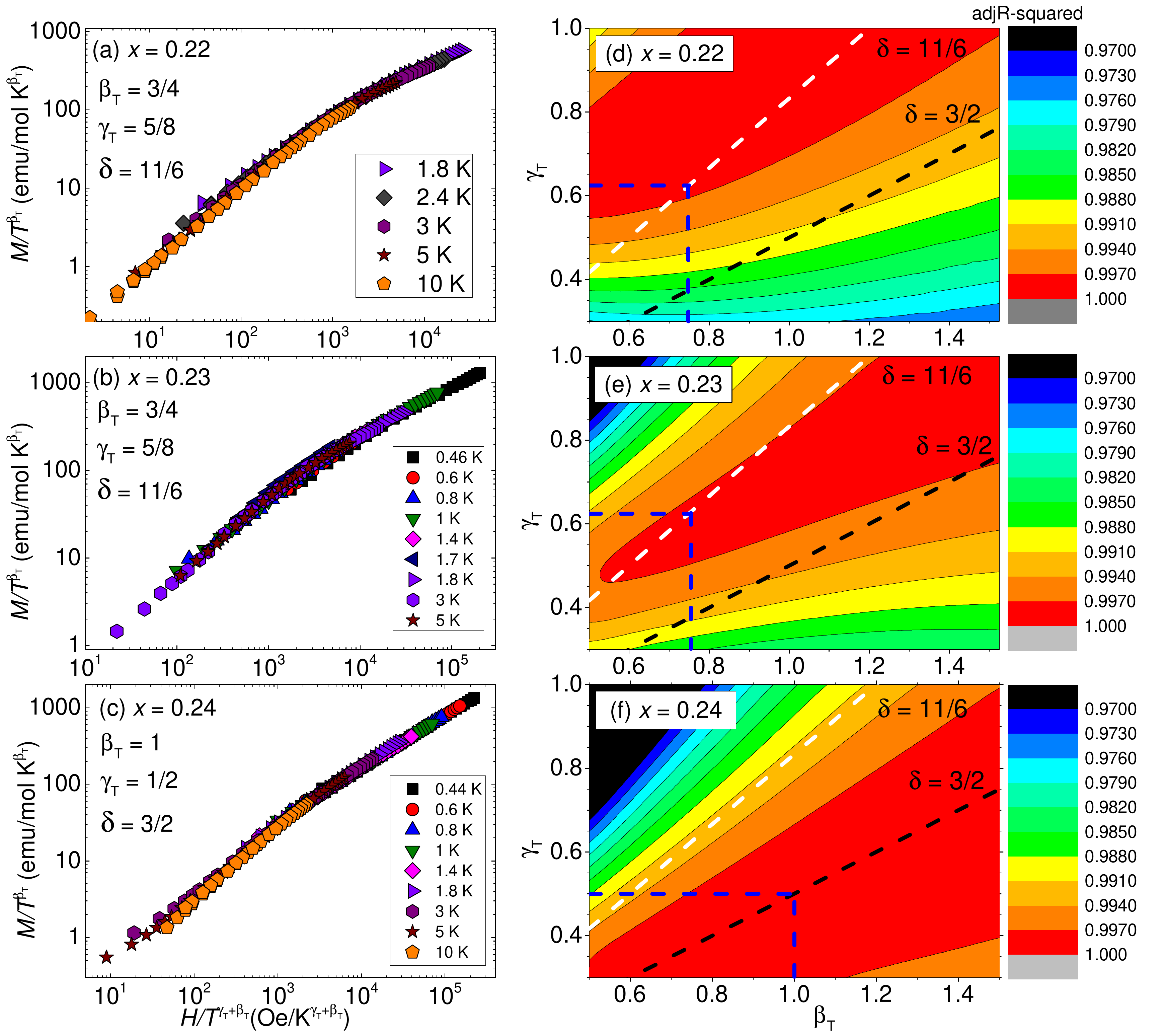}
    \caption{(Color online) Scaling of the magnetization $M$ of \UCFG\ as function of magnetic field $H$ and temperature $T$. (a)-(c) $M/T^{\beta_T}$ vs. $H/T^{\gamma_T + \beta_T}$ for the Fe concentrations $x = 0.22$,  0.23 and 0.24, respectively. The respective critical exponents $\beta_T$, $\gamma_T$, and $\delta$ are denoted in each plot (see main text). (d)-(f) The corresponding adjusted $R^2$ value that describes the goodness of fit for the scaling of $M$ in (a)-(c) ($R^2$ = 1 is the best agreement between data and fit) for a wide range of combinations of the critical exponents ($\beta_T$, $\gamma_T$) for each $x$. The details of how to calculate the adjusted $R^2$ value are provided in the supplemental material~\cite{Supp}. The black and white dashed lines denote the Widom relationship $\gamma_T = \beta_T (\delta -1)$ that relates the critical exponents $\beta_T$ and $\gamma_T$ with the exponent $\delta$ determined from Fig.~\ref{fig:delta}. Here the black and white line use the asymptotic and pre-asymptotic values of $\delta$ respectively (see text and table~\ref{tab:critical}). The blue dashed lines denote the values of ($\beta_T$, $\gamma_T$) used for the modified scaling plots of $M$ in (a)-(c).
    }
    \label{fig:betagamma}
\end{center}
\end{figure}

Because the exponent $\delta$ takes on the value associated with the quantum critical point only for $T=0$, it is crucial to inspect the critical behavior as a function of temperature. Here the critical exponents, $\beta_T$ and $\gamma_T$, describe the temperature-scaling of $M$($T$, $H$), where isotherms converge onto a single curve when displayed as $M$($T$, $H$)$/T^{\beta_T}$ vs. $H/T^{\gamma_T + \beta_T}$. Moreover, $\beta_T$ and $\gamma_T$ are related to $\delta$ via the so-called Widom relationship $\gamma_T = \beta_T (\delta -1)$~\cite{Belitz:2015}.

To confirm our findings for $\delta$ we plot the measured isotherms of the magnetization from Figs.~\ref{fig:delta}(a)-(c) via the scaling relation $M$($T$, $H$)$/T^{\beta_T}$ vs. $H/T^{\gamma_T + \beta_T}$ using the values for $\beta_T$ and $\gamma_T$ predicted by BKV theory for a metal with significant disorder (cf. table~\ref{tab:critical}). Here, according to our analysis for $\delta$, we have used the pre-asymptotic values for $\beta_T$ and $\gamma_T$ for $x=0.22$ and $0.23$, and the asymptotic values for $x=0.24$.  As shown in Fig.~\ref{fig:betagamma}, the scaling works remarkably well for all three concentrations, especially at low temperatures, thus supporting our results for $\delta$, and highlighting the excellent agreement with BKV theory.

\begin{figure}[ht!]
\begin{center}
    \includegraphics[width=0.7\columnwidth]{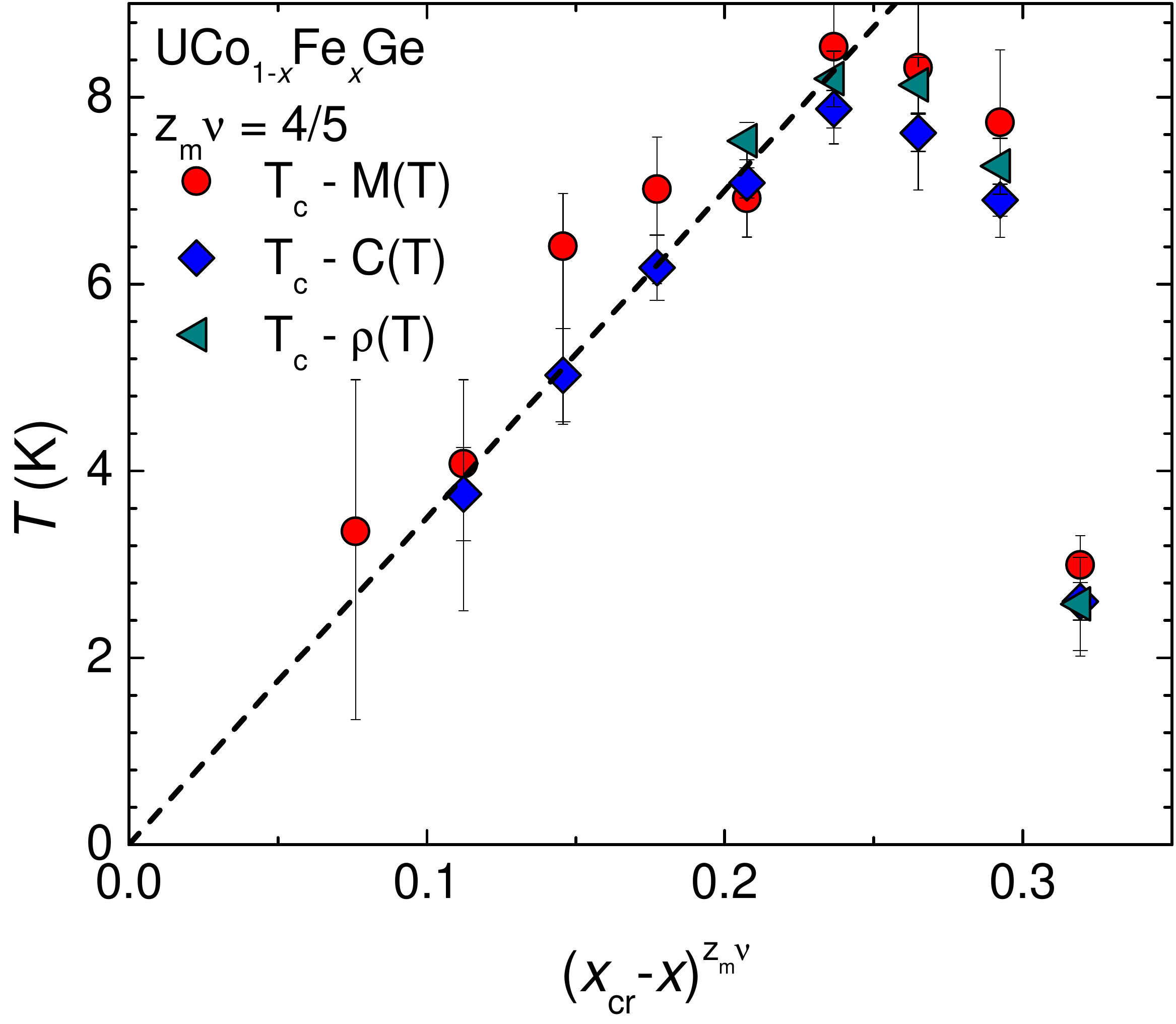}
    \caption{(Color online) The Curie temperature $T_{\rm{C}}$ of \UCFG\ as a function of $(x-x_{cr})$. Here $x$ is the Fe concentration and $x_{cr}$ denotes the concentration at which $T_{\rm{C}}$ is suppressed to zero temperature. The black line is a fit to $T_{\rm{C}} = (x-x_{cr})^{z_m \nu}$ with $z_m \nu=4/5$ for $x_{cr}=0.24$. The red circles, blue diamonds, and green triangles denote $T_{\rm{C}}$ determined from magnetization $M(T)$, specific heat $C(T)$, and electrical resistivity $\rho(T)$ from Ref.~\cite{Huang:2013}.}
    \label{fig:nu}
\end{center}
\end{figure}

We note, however, that detailed analysis of the scaling of $M$($T$, $H$) reveals that the theoretical BKV values of $\beta_T$ and $\gamma_T$ are not the only combination that produces good scaling. This is illustrated in Fig.~\ref{fig:betagamma}(d)-(f), where we plot the adjusted $R^2$ value that describes the goodness of fit ($R^2$ = 1 is the best agreement) for a wide range of combinations ($\beta_T$, $\gamma_T$) (see supplemental material \cite{Supp}). From the small differences in $R^2$ for the various combinations ($\beta_T$, $\gamma_T$), it is clear that the magnetization data are not sensitive enough to pick a single set of values. This is also confirmed by visual inspection of the scaling plots of $M$, where the scaling looks identical for all combinations ($\beta_T$, $\gamma_T$) that correspond to the red areas in Fig.~\ref{fig:betagamma}(d)-(f) (see supplemental material \cite{Supp}). However, a consistent set of values is obtained when the Widom relationship that also takes into account $\delta$ is considered. For reference, we plot the Widom relationship for each concentration in Fig.~\ref{fig:betagamma}(d)-(f), where the dashed black and white lines use the asymptotic and pre-asymptotic value of $\delta$ determined from BKV theory (cf. table~\ref{tab:critical}). From inspection of Fig.~\ref{fig:betagamma}(d) and (e), it is clear that combinations ($\beta_T$, $\gamma_T$) that lead to the best scaling for $x=0.22$ and $0.23$ generally agree better with the pre-asymptotic value of $\delta$ (i.e., the white dashed line runs through the red area with $R^2\approx 1$). In contrast, for $x=0.24$ the asymptotic value of $\delta$ gives better agreement as shown in Fig.~\ref{fig:betagamma}(f) (black dashed line runs through the red area with $R^2\approx 1$) in agreement with our results from Fig.~\ref{fig:delta}.

Fig.~\ref{fig:betagamma}(f) also demonstrates that the asymptotic values ($\beta_T=1$, $\gamma_T=1/2$) (cf. blue dashed lines) lie well in the center of combinations ($\beta_T$, $\gamma_T$) that lead to the excellent scaling of $M$($T$, $H$). This further highlights how well the data for $x=0.24$ agrees with the asymptotic scaling predicted by BKV theory, and suggest that $x=0.24$ is the critical Fe concentration, or is at least very near to it. Similarly, as shown in Fig.~\ref{fig:betagamma}(d) and (e) for $x=0.22$ and $0.23$, the pre-asymptotic values ($\beta_T=3/4$, $\gamma_T=5/8$) (cf. blue dashed lines) belong to the combinations of ($\beta_T$, $\gamma_T$) that converge all isotherms of $M$ onto a single curve.

The product of two additional critical exponents, $\nu$ and $z_m$, can be determined from $M$. Here $z_m$ is the relevant dynamical exponent~\cite{Belitz:2015}. Together they describe how $T_{\rm{C}}$ is suppressed as function of tuning parameter $x$ via $T_{\rm{C}} = (x-x_{cr})^{z_m \nu}$. The theoretical asymptotic and pre-asymptotic values for $z_m \nu$  are 2 and 8/5, respectively (table~\ref{tab:critical}). According to BKV theory, and also in agreement with our data that only shows asymptotic scaling for $x=0.24$, the asymptotic regime typically only exist in a very narrow region around the QPT. Further, as demonstrated above for $x\leq0.23$, the magnetization exhibits pre-asymptotic scaling, suggesting that $T_{\rm{C}}(x)$ should scale with $z_m \nu=8/5$. In contrast, we find that $T_{\rm{C}}(x)$ as determined in Ref.~\cite{Huang:2013} scales well with $z_m \nu=4/5$, as we show in Fig.~\ref{fig:nu}. $z_m \nu=4/5$ is consistent with a Hertz type ferromagnetic QPT in the dirty limit (cf. table~\ref{tab:critical}). Despite the fact that a Hertz's fixed point is unstable, it is expected from BKV theory that it will determine the observable behavior over sizable regions of the phase diagram in many disordered materials \cite{Belitz:2015}. Only in close vicinity of the QPT is it expected to observe the (pre-) asymptotic behavior. However, because the phase transition rapidly becomes extremely broad for $x\longrightarrow x_{cr}$, the scaling of $T_{\rm{C}}(x)$ near $x_{cr}$ cannot be precisely determined.

To summarize, the salient findings of our study are: (i) the critical exponent $\delta$ defined via $M$($T \longrightarrow 0$, $H$)~$\propto H^{1/\delta}$ is equal to 3/2 for the critical Fe concentration $x_{cr}=0.24$ and is significantly smaller than 3/2 for $x=0.22$ and $0.23$ slightly away from the QPT; (ii) our data are in excellent agreement with the critical exponents $\beta_T$ and $\gamma_T$ that describe the temperature-scaling of $M$($T$, $H$) being 1 and 1/2 for $x=0.24$, and 3/4 and 5/8 for $x=0.22$ and $0.23$, respectively; (iii) our results are consistent with the QPT being at or in the very near vicinity of $x_{cr}=0.24$; (iv) the asymptotic and pre-asymptotic regions are situated at Fe concentration $x>0.23$ and $x\leq0.23$, respectively; (v) the $T_{\rm{C}}$ is found to scale as $T_{\rm{C}} = (x-x_{cr})^{z_m \nu}$ for a wide range of $x$ with $z_m \nu=4/5$ consistent with a Hertz's fixed point and in agreement with BKV theory. This establishes that \UCFG\ is the first metal that exhibits critical behavior near a ferromagnetic QPT accessed by chemical substitution that is in complete agreement with the predictions of BKV theory for a metal with significant disorder.

For completeness, we note that a critical exponent $\delta=3/2$ has already been observed in URu$_{2-x}$Re$_x$Si$_2$~\cite{Bauer:05} and Sr$_{1-x}$A$_x$RuO$_3$ \cite{Itoh:2008} at $T$ = 1.8 K  and $T=5$~K, respectively.  However, these results were obtained at finite temperature, whereas the result of BKV theory is only valid for $T \longrightarrow 0$ and in immediate vicinity of the QPT.  Notably, for URu$_{2-x}$Re$_x$Si$_2$, the QPT is situated at $x_c= 0.15-0.2$ but $\delta=3/2$ was obtained for  $x = 0.3$~\cite{Bauer:05, Butch:09}. Finally, all other exponents $\beta_T$, $\gamma_T$, and $z_m \nu$ either do not agree with BKV theory ~\cite{Bauer:05, Butch:09} or were not determined.

In conclusion, our results demonstrate that metals with significant disorder exhibit second-order ferromagnetic QPTs that are more continuous than in Hertz-Millis theory. This establishes for the first time that BKV theory not only describes clean materials extremely well, but is also able to calculate the critical exponents for disordered metals. Our work also identifies several reasons why no other disordered  materials have been reported to show BKV critical exponents. Most notably, it is difficult to achieve the right amount of disorder to observe such behavior; for too small disorder, the tricritical point remains at non-zero temperatures and no quantum critical behavior occurs. In contrast, for too strong disorder, quantum Griffiths effects that compete with the critical behavior are expected~\cite{Brando:2016}. Moreover, there is currently no clear method of determining disorder quantitatively to allow for comparison with theory, and choosing a suitable material is challenging. For example, the often used residual resistivity ratio is only useful in comparing the disorder for different specimens of the same material, and is not meaningful for distinct materials. Further, as shown in detail here, special diligence is required to obtain meaningful critical exponents, as typical magnetization measurements are too insensitive to choose a single set of critical exponents (see supplemental material \cite{Supp}). Finally, for \UCFG, we have shown that the asymptotic scaling is only observed in a tiny region around the QPT~\cite{Kirkpatrick:2014}, which implies that great care is required to identify this region. However, our study may provide a recipe for identifying further disordered metals that exhibit BKV critical exponents near a ferromagnetic QPT.

\begin{acknowledgements}
We are grateful to Dietrich Belitz and Ted Kirkpatrick for useful discussions. Work at Los Alamos National Laboratory (LANL) was performed under the auspices of the U.S. DOE, OBES, Division of Materials Sciences and Engineering. Research at UCSD was supported by the U.S. DOE, BES under Grant No. DE-FG02-04ER46105. K. H. acknowledges financial support through a Seaborg Institute Research Fellowship.

\end{acknowledgements}

\end{bibunit}

\begin{bibunit}
\balancecolsandclearpage
\section*{SUPPLEMENTAL MATERIAL}

\setcounter{section}{0}

In this supplemental material we describe the chemical analysis performed on the samples used in this study as well as the fitting process used to determine $\beta_T$ and $\gamma_T$ from the critical scaling analysis on \UCFG.

\section{Chemical analysis}
\begin{figure*}[h]
\begin{center}
    \includegraphics[width=1\columnwidth]{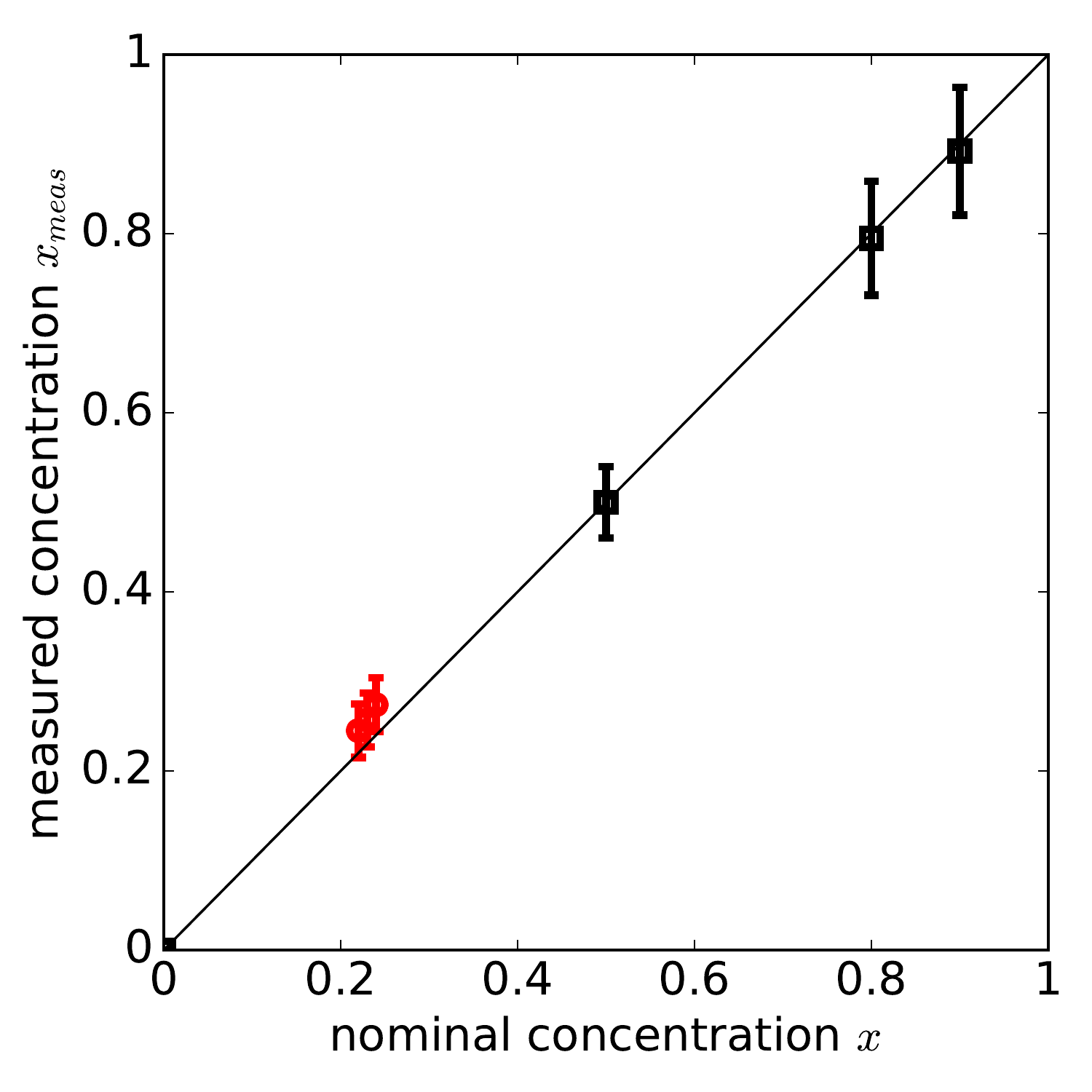}
    \caption{(Color online) The measured Fe concentration $x_{meas}$ as determined via a scanning electron microscope equipped with a energy dispersive spectroscopy microprobe (see text for details) is plotted vs the nominal Fe concentration $x$. The black squares and red circles are measurements carried out for our previous\cite{Huang:2013} and current study, respectively.
    }
    \label{fig:EDS}
\end{center}
\end{figure*}

The same samples of \UCFG\ with Fe concentrations $x = 0.22$, $0.23$, and $0.24$ investigated by magnetization measurements (described in the main text) were characterized by elemental analysis using a commercial scanning electron microscope (FEI Inspect F) equipped with a energy dispersive spectroscopy (EDS) microprobe. From the EDS analysis, we have extracted the actual $x_{meas}$ concentration for each nominal concentration $x$. Each sample was measured at 5 or 6 different positions where we found that within the error bars of this method all measurements for a single sample agree, suggesting that Fe substitution is homogeneous. The results of our analysis are shown in Fig.~\ref{fig:EDS}. In addition to the samples measured for this study (red circles) we have added measurements for samples from our previous study Ref.~\cite{Huang:2013} in Fig.~\ref{fig:EDS} (black squares). Together these results establish that the nominal Fe concentration agrees with the actual concentration within the error bars throughout the entire series. Therefore, we use the nominal Fe concentration $x$ throughout both the supplemental material and the main text.

\section{Determination of critical exponents $\beta_T$ and $\gamma_T$}

\begin{figure*}[h]
\begin{center}
    \includegraphics[width=1\columnwidth]{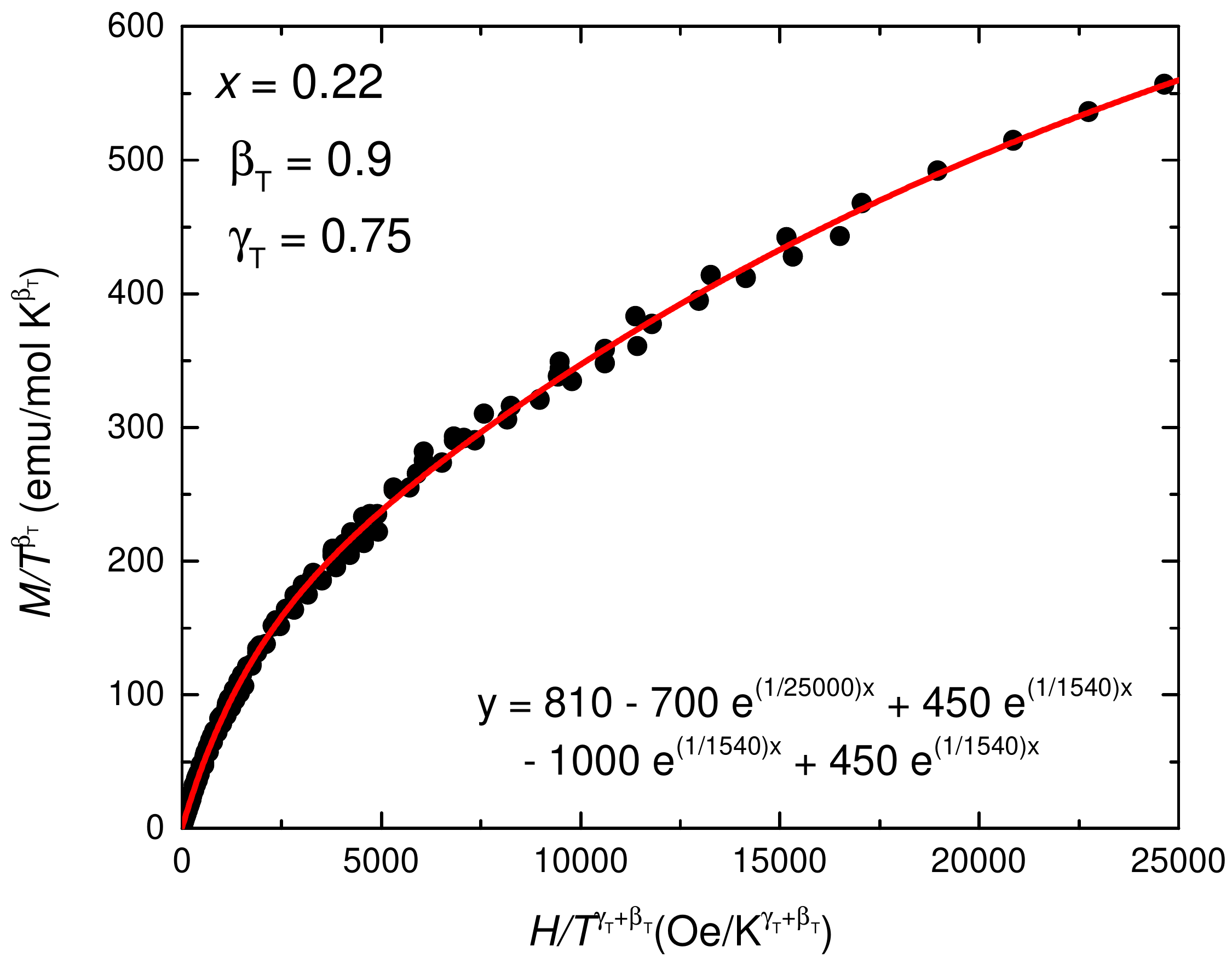}
    \caption{(Color online) Exemplary fit of the magnetization $M$ of \UCFG\ with the Fe concentration $x=0.22$ to Eq.~\ref{seq1}, where $\beta_T=0.9$ and $\gamma_T=0.75$ were selected.
    }
    \label{fig:fit}
\end{center}
\end{figure*}

\begin{figure*}[h]
\begin{center}
    \includegraphics[width=1.9\columnwidth]{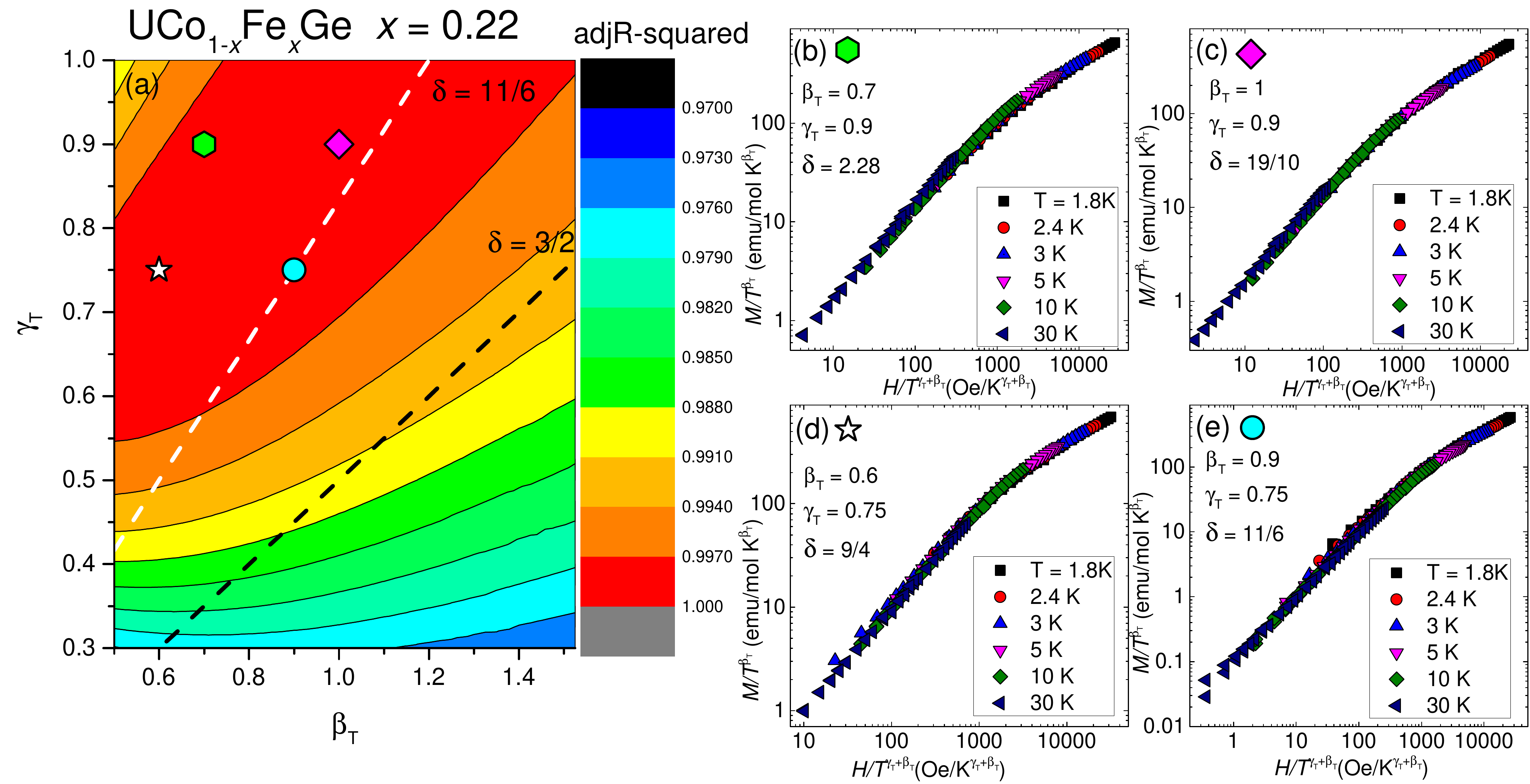}
    \caption{(Color online) (a) The adjusted $R^2$ value that describes the goodness of fit for fitting the magnetization $M$ of \UCFG\ with the Fe concentration $x=0.22$ to Eq.~\ref{seq1} for a wide range of combinations of the critical exponents ($\beta_T$, $\gamma_T$) ($R^2$ = 1 is best agreement between data and fit). Here both $\beta_T$ and $\gamma_T$ were varied from 0.3 up to 1.5 in 0.025 increments. The black and white dashed lines denote the Widom relationship $\gamma_T = \beta_T (\delta -1)$ that relates the critical exponents $\beta_T$ and $\gamma_T$ with the exponent $\delta$ (see main text). Here the black and white line use the asymptotic and pre-asymptotic values of $\delta=3/2$  and $\delta=11/6$, respectively. Displayed in panels (b)-(e) are the magnetization data plotted as $M$($T$, $H$)$/T^{\beta_T}$ vs. $H/T^{\gamma_T + \beta_T}$ for representative combinations of $\beta_T$ and $\gamma_T$ in the red region, which represents values  values that result in a good fit. The colored symbols (green hexagon, purple diamond, white star, and cyan circle) in the top left corner of panels (b)-(e) represent their location on the plot in panel a according to the $\beta_T$ and $\gamma_T$ values.}
    \label{fig:red}
\end{center}
\end{figure*}

\begin{figure*}[h]
\begin{center}
    \includegraphics[width=1.9\columnwidth]{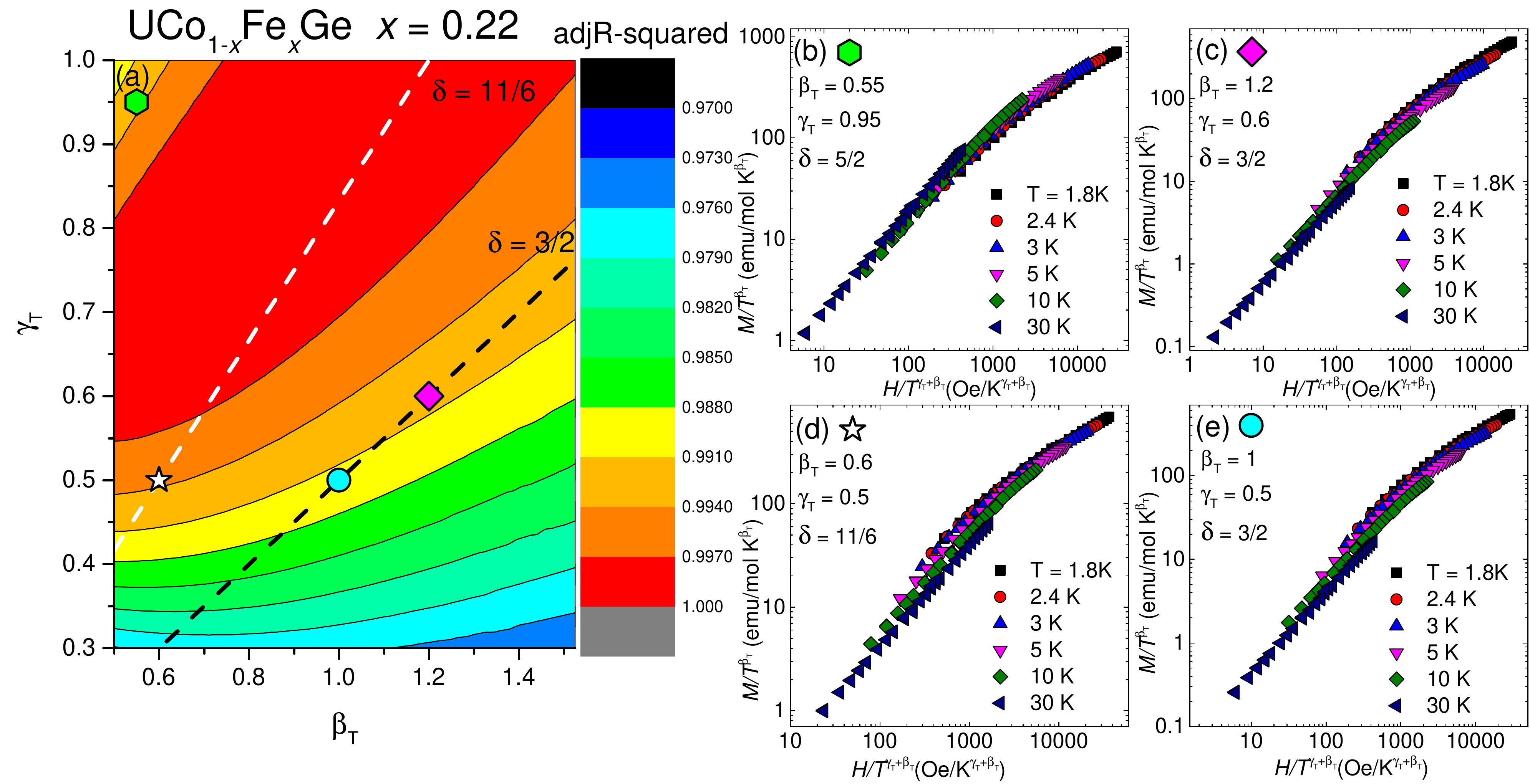}
    \caption{(Color online) (a) The adjusted $R^2$ value that describes the goodness of fit for fitting the magnetization $M$ of \UCFG\ with the Fe concentration $x=0.22$ to Eq.~\ref{seq1} for a wide range of combinations of the critical exponents ($\beta_T$, $\gamma_T$) ($R^2$ = 1 is best agreement between data and fit). Here both $\beta_T$ and $\gamma_T$ were varied from 0.3 up to 1.5 in 0.025 increments. The black and white dashed lines denote the Widom relationship $\gamma_T = \beta_T (\delta -1)$ that relates the critical exponents $\beta_T$ and $\gamma_T$ with the exponent $\delta$ (see main text). Here the black and white line use the asymptotic and pre-asymptotic values of $\delta=3/2$  and $\delta=11/6$, respectively. Similar as in Fig.~\ref{fig:red} panels (b)-(e) show the magnetization data plotted as $M$($T$, $H$)$/T^{\beta_T}$ vs. $H/T^{\gamma_T + \beta_T}$ for representative combinations of $\beta_T$ and $\gamma_T$. However, here we plot combinations of $\beta_T$ and $\gamma_T$ outside of the red region that represents values that result in a good fit. The colored symbols (green hexagon, purple diamond, white star, and cyan circle) in the top left corner of panels (b)-(e) represent their location on the plot in panel a according to the $\beta_T$ and $\gamma_T$ values.}
    \label{fig:notred}
\end{center}
\end{figure*}

As described in the main text, the critical exponents, $\beta_T$ and $\gamma_T$, describe the temperature-scaling of $M$($T$, $H$), where isotherms converge onto a single curve when displayed as $M$($T$, $H$)$/T^{\beta_T}$ vs. $H/T^{\gamma_T + \beta_T}$ when the correct set of ($\beta_T$, $\gamma_T$) is chosen. However, when relying on a pure visual inspection of the scaling, it becomes apparent that many sets of ($\beta_T$, $\gamma_T$) yield scaling that is qualitatively similar. To quantify the quality of the temperature-scaling of the magnetization for the various combinations of $\beta_T$ and $\gamma_T$ , we followed a procedure that has recently been established for Sr$_{1-x}$Ca$_x$RuO$_3$.\cite{Huang:2015} and is described in the following. Our magnetization data for \UCFG\ were fitted to the following equation:

\begin{equation}\label{seq1}
y = A_0 + \sum_{n = 1}^{4} A_n e^{(1-T_n) x},
\end{equation}

where $y$ = $M$($T$, $H$)$/T^{\beta_T}$, $x$ = $H/T^{\gamma_T + \beta_T}$, and both $A_n$ and $T_n$ are fitting parameters. Fig.~\ref{fig:fit} shows a representative fit for the Fe concentration $x = 0.22$ with $\beta_T$ = 0.75 and $\gamma_T$ = 0.6. For each concentration of $x$, a wide range of $\beta_T$ and $\gamma_T$ values were analyzed, from 0.3 up to 1.5 in increments of 0.025 with the results displayed in Fig.~\ref{fig:red}(a). The different colored regions correspond to the quality of fit, where the red region represents the best fit parameters for which the adjusted $R^2$ values are closest to 1. Displayed in Figs.~\ref{fig:red}(b)-(e) are the magnetization data for different combinations of $\beta_T$ and $\gamma_T$. Here the symbols plotted in the upper left corner of each subfigure mark the position in panel Fig.~\ref{fig:red}(a). Similarly, in Fig.~\ref{fig:notred} we show various fits with combinations of ($\beta_T$, $\gamma_T$) that yield lower quality fits and width adjusted $R^2$ values that lay outside the red region in Fig.~\ref{fig:notred}(a). Finally, $\beta_T$ and $\gamma_T$ are related to $\delta$ via the so-called Widom relationship $\gamma_T = \beta_T (\delta -1)$~\cite{Belitz:2015} that is plotted in Fig.~\ref{fig:red}(a) and Fig.~\ref{fig:notred}(a). Here the black and white lines use the asymptotic and pre-asymptotic values of $\delta$ respectively (see table 1 in main text for details and Ref.~\cite{Belitz:2015}).

\begin{acknowledgments}

Work at Los Alamos National Laboratory (LANL) was performed under the auspices of the U.S. DOE, OBES, Division of Materials Sciences and Engineering and funded in part by the LANL Directed Research and Development program. Research at UCSD was supported by the U.S. DOE, BES under Grant No. DE-FG02-04ER46105. K. H. acknowledges financial support through a Seaborg Institute Research Fellowship.

\end{acknowledgments}

\end{bibunit}

\end{document}